\documentclass{article}
\usepackage{graphics}
\usepackage[english]{babel}
\usepackage[utf8x]{inputenc}
\usepackage[per-mode=symbol]{siunitx}
\usepackage{amsmath}
\usepackage{amssymb}

\usepackage{authblk}
\usepackage{graphicx}
\usepackage{dcolumn}
\usepackage{bm}
\usepackage{vmargin}
\setpapersize{A4}
\setmargins{2.5cm}
{1.5cm}                        
{16.5cm}                      
{23.42cm}                    
{10pt}                           
{1cm}                           
{0pt}                             
{2cm}

\usepackage{graphicx}
\usepackage{dcolumn}
\usepackage{bm}

\usepackage{textcomp}

\title{
Site-bond percolation solution to preventing the propagation of \textit{Phytophthora} zoospores on plantations
}

\author[1,2]{J. E. Ram\'irez\thanks{Corresponding Author: jerc.fis@gmail.com}}
\author[1,2]{C. Pajares}
\author[3]{M. I. Mart\'inez}
\author[2]{R. Rodr\'iguez Fern\'andez}
\author[4]{E. Molina-Gayosso}
\author[4]{J. Lozada-Lechuga}
\author[3]{A. Fern\'andez T\'ellez}
\affil[1]{Departamento de F\'isica de Part\'iculas, Universidad de Santiago de Compostela, E-15782 Santiago de Compostela, Espa\~na}
\affil[2]{
Instituto Galego de F\'isica de Altas Enerx\'ias, Universidad de Santiago de Compostela, E-15782 Santiago de Compostela, Espa\~na}
\affil[3]{Facultad de Ciencias F\'isico Matem\'aticas, Benem\'erita Universidad Aut\'onoma de Puebla, Apartado Postal 165, 72000 Puebla, Pue., M\'exico}
\affil[4]{Universidad Polit\'ecnica de Puebla, Tercer carril del Ejido Serrano, 72640, Juan C. Bonilla, Pue., M\'exico}

\date{}

\begin{document}
\maketitle
\begin{abstract}
We propose a strategy based on the site-bond percolation to minimize the propagation of \textit{Phytophthora} zoospores on plantations, consisting in introducing physical barriers between neighboring plants. Two clustering processes are distinguished: i) one of cells with the presence of the pathogen, detected on soil analysis; and ii) that of diseased plants, revealed from a visual inspection of the plantation. The former is well described by the standard site-bond percolation. In the latter, the percolation threshold is fitted by a Tsallis distribution when no barriers are introduced. We provide, for both cases, the formulae for the minimal barrier density to prevent the emergence of the spanning cluster.
Though this work is focused on a specific pathogen, the model presented here can also be applied to prevent the spreading of other pathogens that disseminate, by other means, from one plant to the neighboring ones.
Finally, the application of this strategy to three types of commercialy important Mexican chili plants is also shown.
\end{abstract}


\section{Introduction}
The genus \textit{Phytophthora} (from Greek, meaning \textit{phyto}, ``plant,'' and \textit{phthora}, ``destroyer'' \cite{bio1,bio2,bio3})
is one of the most aggressive phytopathogens that attack the roots of plants and trees in every corner of the world.
The diseases caused by exposition to \textit{Phytophthora} generate tremendous economical losses in agronomy and forestry.
For example, \textit{P. capsici} cause considerable damage in plantations of chili, cucumber, zucchini, etc.~\cite{chiles3,chiles,chiles2}.
The same occurs with tomato and potato plantations, which are affected by \textit{P. infestants} \cite{papa,papa2, papa3}.
\textit{P. cinnamomi} harms avocado plantations \cite{aguacate3,aguacate,aguacate2} and, together with \textit{P. cambivora}, produce the ink disease which is widely distributed along Europe \cite{ink1,ink2,ink3}.
\textit{Phytophthora} has caused significant devastation on Galician chestnut and the Australian eucalypt, putting them close to extinction \cite{castano,castano2,castano3}.

From a biological perspective, \textit{Phytophthora} shares morphological characteristics with true fungi (Eumycota) such as mycelial growth or the dispersion of spores of mitotic or asexual origin. Its form of locomotion, by means of flagella \cite{bio6}, is a distinctive feature that enables them to have a great impact on the plant kingdom as phytopathogens. They can disperse through soil moisture or water films including those on the surface of the plants.
These motile zoospores, emerging from mature sporangia in quantities of 20 to 40, can swim chemotactically towards the plants \cite{bio6,bio7,bio8}. When they reach the surface of the roots they lose their flagella, encyst in the host and form a germination tube through which they penetrate the surface of the plant \cite{bio9,bio10}. Moreover, many species of \textit{Phytophthora} can persist as saprophytes if the environmental conditions are not appropriate, but become parasitic in the presence of susceptible hosts \cite{bio8}. Due to the physiology of the oomycetes most of the fungicides have no effect on them \cite{bio1,fungi3, fungi2, *fungi1}.
Therefore, research on non-chemical strategies that minimize or eliminate the propagation of the pathogen is necessary.

It has been noticed that for some type of plants not all individuals manifest the disease after the exposition to a specific pathogen.
We take advantage of this fact to define the pathogen susceptibility ($\chi$) of a plant type as the fraction of individuals that get the disease. It can be interpreted as the probability that a sample of the plant gets sick after being exposed to the pathogen,
and can be measured in a laboratory or a greenhouse under controlled conditions, or by direct observation in the plantation.

On the other hand, one of the models widely used to describe physical processes is the site-bond percolation, that has been applied to study the spread of diseases \cite{sp,sp4,sp2,sp3}.
It is a generalization of the site and bond percolation that consists in determining both site and bond occupation probabilities needed to the emergence of a spanning cluster of sites connected by bonds. In this context, two nearest-neighboring sites do not belong to the same cluster if there is not a bond connecting them.
In this work, occupied sites in the percolation system represent susceptible plants through which the propagation process can occur, and bonds represent the direction of propagation of the pathogen.

It is worth mentioning that zoospores move directly to neighboring plants. Placing physical barriers between them (that is, perpendicularly to the direction of propagation) can help to decrease the opportunity for root to root pathogen transmission.
For instance, the Australian government recommends using physical root barriers such as impermeable membranes made of high-density polyethylene~\cite{barrier1,barrier2,barrier3,barrier4}, which have been used in agriculture and horticulture.
Trenches filled with compost (a mixture of manure and crop residues) in addition with biological control agents (for example \textit{Trichoderma spp.} or \textit{Bacillus spp.}) could be used as a good barrier against soil-borne pathogens like oomycetes and fungi \cite{edu1,edu2,edu3}.
With the 
use of 
barriers it could be possible to fragment the spanning cluster of susceptible plants, preventing the propagation of the pathogen.
Thus, if the pathogen susceptibility of the plant is known, one can try to determine the minimal density of barriers ($p_\text{w}$) that stops the propagation of the pathogen. However, this density does not necessarily corresponds to the bond percolation threshold.

Although this paper is motivated by the important problem caused by the propagation of \textit{Phytophthora}, which is still unsolved nowadays, the strategy presented here can be adapted to mitigate the spread of other diseases. There exist other phytopathogens relevant to agronomy that disseminate over neighboring plants by, for example, walking \cite{walk}, rain splashing \cite{splash,splash3,splash2}, swimming \cite{pythium}, etc.; such as the red spider mites, leaf rust, \textit{Pythium} (with similar propagation mechanisms as \textit{Phytophthora}) among others. In practice one only needs to find a suitable physical barrier that efficiently avoids nearest-neighbor propagation of the specific phytopathogen.

In Sec.~\ref{model}, we introduce the site-bond percolation model for the pathogen-plant interaction and the role of the barriers. Section~\ref{simulation} describes the simulation method used in this work and provides the simulation rules for the clustering process. It also shows an example of the simulation process and describes the data analysis method.
In Sec.~\ref{results}, we report the critical curves as a function of the initial percentage of inoculated soil for the barrier-free case. These curves indicate the maximum value of the pathogen susceptibility that guarantees a spanning cluster of diseased plants is not formed even if the soil is completely infested with the pathogen. Additionally, we provide the empirical formulae to determine the density of barriers that prevents the emergence of the spanning cluster when the susceptibility exceeds the aforementioned critical value.
In Sec.~\ref{chili}, we show the application of this method to three varieties of Mexican chili plants with high comercial value.
Finally, Section~\ref{conclusions} presents the conclusions of this work.

\begin{figure}
\centering
\includegraphics[scale=0.33]{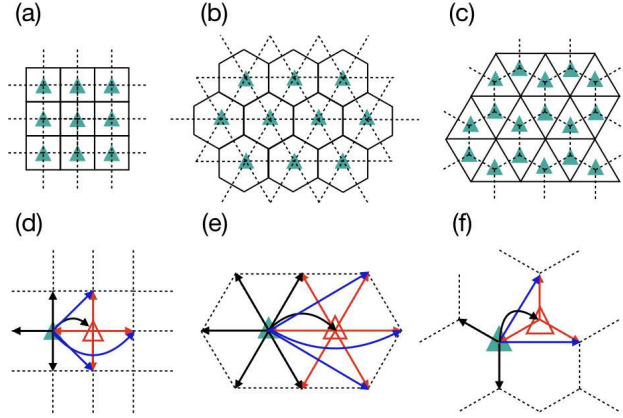}
\caption{Possible barrier locations (solid lines), directions of microorganism propagation (dotted lines), and modification of the nearest neighbor meaning induced by inoculated cells with a resistant plant in square [(a) and (d)], triangular [(b) and (e)], and honeycomb [(c) and (f)] lattices. Bottom figures show susceptible plants (green triangle) with a neighboring resistant plant in an inoculated cell ( red triangle). As a consequence of the microorganism propagation (red arrows), the nearest neighbor definition (black arrows) is modified since the site with the susceptible plant can now be linked to farther sites (blue arrows).}
\label{fig:lattices}
\end{figure}

\section{Model} \label{model}
The plantation is modeled as a simple two-dimensional lattice (square, triangular, and honeycomb) wherein each site represents a plant.
The lattice spacing is chosen as the maximum displacement length that the pathogen can travel before entering a state of dormancy or before dying due to starvation.
This condition ensures the pathogen can only move to the nearest neighbor cells as depicted in Fig.~\ref{fig:lattices}.
We assume a site with an active pathogen will propagate the disease to all nearest-neighbor sites.

Here, the pathogen susceptibility plays an important role since resistant plants can act as a natural barrier for susceptible plants by locally containing the propagation process, i. e. a resistant plant does not disseminate the disease. In our model resistant plants are uniformly distributed on the system since it is not possible to determine in advance which seeds will grow into resistant or susceptible plants.
In this way the pathogen susceptibility plays the role of the occupation probability in the traditional treatment of percolation theory.

Another essential variable that needs to be considered is the initial fraction of inoculated cells at the beginning of the propagation process which is denoted by $I$. In our model these cells are distributed uniformly over the lattice. This parameter is relevant to amalgamate adjacent-disjoint clusters promoting a favorable environment for the formation of a spanning cluster of diseased plants or of cells with the presence of the pathogen \cite{jerc}.
Additionally, we put barriers that are randomly distributed in the lattice. These are placed perpendicularly to the direction of propagation of the pathogen (see Fig.~\ref{fig:lattices}), and its primary function is to prevent the pathogen from reaching neighbor sites.
Note that all possible barriers that can be placed form the dual lattice to that formed by all possible directions of propagation of the pathogen.
Then the question we want to answer is: what is the minimal barrier density, in terms of $\chi$ and $I$, that guarantees a spanning cluster will not appear?

We distinguish two different clustering processes: i) the formation of clusters of cells with the presence of the pathogen, and ii) the formation of clusters of diseased plants. Although both processes are consequence of the propagation of the pathogen they depend in different ways on the intrinsic properties of the plants. In practice one would observe the first process if a pathogen soil test is performed while a visual inspection of the damage on the plantation would reveal the second process. In the following we refer to them as \emph{soil} and \emph{plant} cases respectively, and the corresponding variables will be labeled with a superscript.

In the soil case, for a lattice with $N$ sites, the mean number of available plants $\langle N \rangle_\text{av}$ to the propagation process is $\langle N \rangle_\text{av}=N\chi$. Since the susceptibility of the plant and the inoculation state of the cell are independent variables, it is necessary to take into account the mean number of inoculated cells $\langle N \rangle_\text{in}$ with a resistant plant. This condition adds $\langle N \rangle_\text{in}=N(1-\chi)I$ extra available cells. Thus, the total mean number of cells where the propagation process can occur is $\langle N \rangle_\text{tot}=\langle N \rangle_\text{av}+\langle N \rangle_\text{in}$. Therefore, the propagation takes place in a percolating system with an effective occupation probability $p_\textbf{eff}^\text{soil}=I+(1-I)\chi$.
In this case, the spanning cluster emerges if $p^\text{soil}_\text{eff}\geq p_\text{cs}$, where $p_\text{cs}$ is the critical probability in the purely site percolation. Thus the desired percolation threshold is $p^\text{soil}_\text{eff}= p_\text{cs}$.

The introduction of barriers in the soil case makes the system suitable to be modeled with the site-bond percolation.
The critical curves as a function of the occupation probabilities of sites ($p_\text{s}$) and bonds ($p_\text{b}$) has been empirically fitted using \cite{tarasevich} $p_\text{b}=B/(p_\text{s}+A)$, where $A=(p_\text{cb}-p_\text{cs})/(1-p_\text{cb})$, $B=p_\text{cb}(1-p_\text{cs})/(1-p_\text{cb})$, and $p_\text{cb}$ is the critical probability in the purely bond percolation.
Moreover, since barriers are located in the dual lattice,
the density of barriers and the bond occupation probability are related by $p_\text{b}+p_\text{w}^\text{soil}=1$, that is, the joint-set of barriers and bonds it is exactly ${\mathcal{N}_b}$. So we finally find that the critical curves for the soil case can be written as
\begin{equation}
p^\text{soil}_\text{w}=1-\frac{p_\text{cb}(1-p_\text{cs})}{(1-p_\text{cb})(I+(1-I)\chi)+p_\text{cb}-p_\text{cs}}.
\label{eq:pw-soil}
\end{equation}

On the other hand, for the plant case, inoculated cells with a resistant plant do not belong to the cluster of diseased plants. However, these cells play an essential role since adjacent-disjoint clusters can be amalgamated through them. This fact modifies the nearest neighbor meaning since it is then possible to link two susceptible plants separated by a distance greater than the lattice spacing (see Fig.~\ref{fig:lattices}), then the possibility to amalgamate adjacent-disjoint clusters is increased \cite{jerc}.

The main difference between the soil and plant cases is just this amalgamating role played by inoculated cells with a resistant plant at the beginning of the propagation process. In the soil case, these cells are considered as occupied sites, while in the plant case, they do not belong to any cluster; however, they can transmit the disease over neighboring susceptible plants. Schematically, this latter situation looks like a healthy plant with sick neighbors.

\section{Simulation method}\label{simulation}
We implemented a modified version of the Newman-Ziff algorithm reported in Refs.~\cite{Ziff1,Ziff2} to determine the percolation threshold.

Since the susceptibility condition of each plant and the cells' inoculation state are independent of each other they are stored in separate matrices in the simulation. These matrices, that we call  $\mathbb{X}$ and $\mathbb{I}$, respectively, are initially null. They are then filled according to the predefined values of  $\chi$ and $I$. For the case with no barriers, however, only the knowledge of the inoculated cells is required to determine the percolation thresholds.

For simplicity we describe the implementation of the algorithm for a square lattice. However, this algorithm can also be used for other lattices simply changing the implementation of the nearest neighbor definition.

Each cell of the $L\times L$ matrices $\mathbb{X}$ and $\mathbb{I}$ is labeled with a progressive number $M=iL+j$, for the cell at row $i$ and column $j$. The set of cells' labels is then $\mathcal{N}=\{0,1,2,\dots,L^2-1\}$. On the other side, the possible propagation directions for all cells form a network with $2L(L-1)$ bonds since the system is considered as free of periodic boundary conditions. As we did with the cells, each bond is labeled with progressive numbers that form the set $\mathcal{N}_b=\{0,1,2,\dots,2L(L-1)-1\}$.

An initial number of inoculated cells $n_I$ is drawn from the binomial distribution $B(L^2,I)$ and then $n_I$ labels are randomly taken from the set $\mathcal{N}$. The corresponding cells are the sites from which the infection process will propagate. These cells are marked  by changing their state from 0 to 1. The initial distribution of susceptible plants, that is plants that will get the disease if they are exposed to the pathogen, is obtained in a similar way. Note that only the initial conditions are set so far and the propagation process has not been started so that no cells are linked yet.

To add bonds between cells the $\mathcal{N}_b$ labels are randomly permuted and then the corresponding bonds are added one at a time until a spanning cluster is formed. It should be recalled that bonds determine the direction of propagation in this model.

To decide which bonds will connect the sites we impose rules based on the way the pathogen transports itself from site to site. Since the zoospores are capable of detecting the presence of neighboring plants, they will swim towards them as soon as they emerge from the sporangia. If a zoospore reaches a resistant plant it will either enter a latency state or die from inanition so that it won't be able to further propagate the disease. If, on the other hand, the zoospore arrives at a susceptible plant, it will attack the plant and produce new sporangia. They, in turn, will produce new zoospores that will eventually swim towards neighboring plants. Thus the rules can be stated as follows.

\noindent A bond will connect two nearest-neighbor sites if:
\begin{enumerate}
\item Soil case:
    \begin{enumerate}
    \item Any of the sites was inoculated during the initial configuration.
    \item Both sites have susceptible plants.
    \end{enumerate}
\item Plant case:
    \begin{enumerate}
    \item Any of the sites was inoculated during the initial configuration and the other has a susceptible plant.
    \item Both sites have susceptible plants.
    \end{enumerate}
\end{enumerate}

This way bonds are added one by one, and sites are connected according to the rules above, until a cluster that connects one side of the lattice to the opposite one, the so-called spanning cluster, appears.
The union-find algorithm is used to connect sites. Since not every site pair can interact not every bond can connect adjacent sites. In order to identify the spanning cluster, before starting the simulation process, susceptible plants in the last and first rows are united with auxiliary labels -1 and -2, respectively. Then, the simulation process is stopped when the labels \{-1,-2\} change to the same value.

The essential difference between the two cases is the role played by the inoculated cells with a resistant plant. In the soil case they become occupied sites while in the plant case they may merge disjoint clusters.

\begin{figure}
\centering
\includegraphics[scale=1]{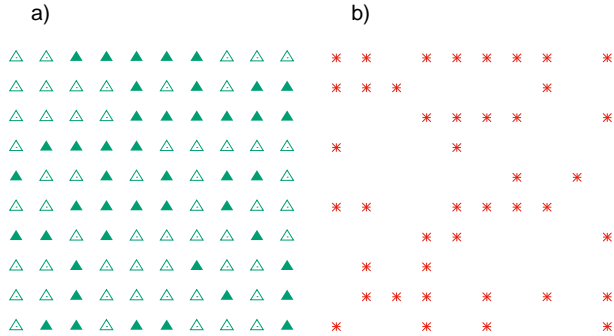}
\caption{
Examples of possible initial configurations of a system of size $L=10$. (a) Distribution of cells with susceptible (filled triangles) and resistant (empty triangles) plants. (b) Distribution of inoculated cells.}
\label{fig:ini}
\end{figure}

To visualize the difference between both cases consider an $L=10$ system with $\chi=0.5$ and $I=0.4$. Figure~\ref{fig:ini} shows one possible initial configuration of susceptible plants and inoculated cells before the propagation process starts.

In a system of size $L=10$ there are 180 bonds. A possible random permutation of their labels is listed below:

\{118, 63, 26, 119, 160, 22, 64, 142, 156, 126, 8, 152, 73, 127, 32, 78, 81, 170, 36, 92, 89, 123, 57, 68, 12, 33, 24, 129, 158, 46, 169, 82, 48, 147, 69, 38, 18, 56, 168, 178, 179, 164, 114, 6, 79, 42, 86, 41, 13, 52, 165, 115, 43, 85, 172, 116, 133, 11, 27, 139, 29, 15, 0, 138, 122, 40, 7, 148, 74, 71, 113, 177, 111, 135, 37, 51, 67, 9, 121, 98, 99, 35, 49, 108, 151, 53, 173, 39, 1, 5, 2, 153, 45, 146, 76, 59, 145, 143, 163, 96, 16, 104, 101, 61, 144, 28, 102, 17, 88, 31, 3, 141, 109, 77, 65, 80, 166, 106, 167, 117, 70, 130, 21, 83, 140, 20, 157, 10, 136, 161, 137, 107, 100, 150, 110, 91, 132, 128, 112, 93, 44, 149, 19, 94, 131, 154, 155, 30, 62, 171, 23, 34, 55, 4, 54, 176, 58, 75, 174, 50, 60, 125, 47, 25, 103, 134, 120, 159, 90, 84, 14, 87, 175, 124, 95, 105, 66, 72, 97, 162\}.

The bonds are added in this order until a spanning cluster appears. The entries of one of the cells a given bond can connect are given by $i=\lfloor h/(2L-1)\rfloor$ and $j=h\mod (2L-1)$, where $h$ is the bond's label and $\lfloor x\rfloor$ denotes the integer part of $x$. Note that the orientation of the bond is identified as horizontal if $j<L-2$ or vertical otherwise. In addition, the value of $j$ should be corrected for vertical bonds by subtracting $L-1$. Then, the cells with entries $i, j$ and $i,j+1$ are taken if the bond is horizontal; while the cells at $i, j$ and $i+1,j$ are taken if the bond is vertical. Finally, if the pair taken fulfills the rules given previously they are connected using the union-find algorithm.

\begin{figure}
\centering
\includegraphics[scale=1]{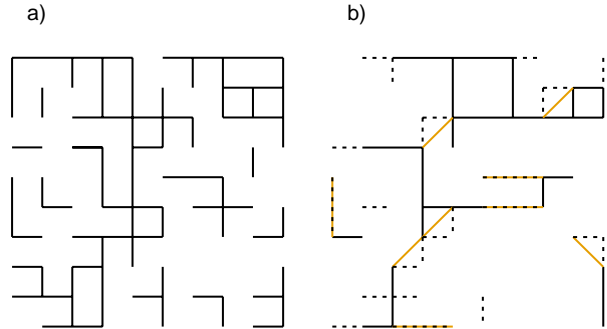}
\caption{
Spanning clusters formed in the a) soil and b) plant cases for the initial conditions of Fig.~\ref{fig:ini} and the list of bonds given in the text. Only the bonds that connect sites are shown (black lines). In the case (b), bonds connecting a resistant plant in an inoculated site to a susceptible plant are represented with dashed lines. Yellow lines show the modification of the nearest neighbor definition.}
\label{fig:bejem}
\end{figure}

Figure~\ref{fig:bejem} shows the networks formed by connected bonds in both cases. While in the soil case 121 bonds were added before the spanning cluster appeared, in the plant case were needed 160 bonds. Note that, although each network has its own topology, in the plant case the fundamental role for the formation of a spanning cluster is played by the modification of the nearest neighbor definition (yellow lines in Fig.~\ref{fig:bejem}b) introduced by the interactions between susceptible plants and inoculated cells with a resistant plant on it (dashed lines in Fig.~\ref{fig:bejem}b). This clearly shows the consequence of this type of interactions, namely their capacity to merge disjoint clusters of susceptible plants.

\subsection{Data analysis}

Using this method, we determined the probability $P_n$ that a spanning cluster appears after adding $n$ bonds (or sites)~\cite{Mertens} as an average over $10^4$ runs for each pair $(\chi,I)$. Starting in $\chi=1$ and $I=1$ we decreased their values independently in steps of $\Delta \chi=\Delta I=0.05$. 
Then the percolation probability is computed as $P(p)=\sum_n  B(N,n,p) P_n$ where $B(N,n,p)$ is the binomial distribution~\cite{Ziff1,Ziff2}, $N$ is the total number of sites or bonds in the lattice and $p$ is the occupation probability of sites or bonds correspondingly.
Lastly, the percolation threshold is determined by solving the equation $P(p_\text{c})=0.5$~\cite{Rintoul_1997}.
To this end, the percolation probability is computed from $\langle n_c \rangle/L^2-0.15$ to $\langle n_c \rangle/L^2+0.15$ in steps of $\Delta p=0.01$. Then, $P(p)=0.5[1+\tanh((p-p_\text{c})/\Delta_L)]$ is fitted to the estimated data. Here $p_c$ is the estimation of the percolation threshold and  $\Delta_L$ is the width of the sigmoid transition \cite{Rintoul_1997}.

To take finite size effects into account we also performed simulations using the system size $L=$ 32, 64, 128, and 256. Thus the percolation threshold in the thermodynamic limit is estimated by the extrapolation of the scaling relation $p_\text{c}-p_\text{c}(L)\propto L^{-1/\nu}$, where $\nu$ is the exponent corresponding to the correlation length
\cite{stauffer}. It is well known that the transition width $\Delta_L$ scales as a function of the system size $L$ as $\Delta_L\propto L^{-1/\nu}$ \cite{Coniglio}. From the fit of the percolation probability data, we found that $\nu=4/3$, which is in good agreement with the results reported in the literature for the percolation theory in 2D.
Finally, the critical density of barriers is calculated as $p_\text{w}=1-p_\text{cb}^*$, where $p_\text{cb}^*$ is the bond percolation threshold as a function of $\chi$ and $I$.

\begin{figure}
\centering
\includegraphics[scale=1]{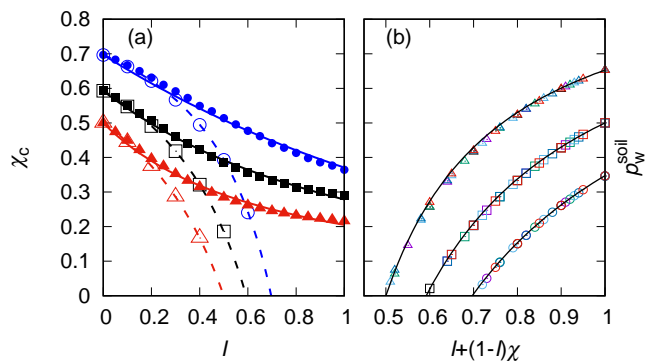}
\caption{(a) Critical curves for cluster formation over infested soil (hollow figures) and infected plants (solid figures) on triangular (triangles), square (squares) and honeycomb (circles) lattices with no barriers. Theoretical curves for the soil case (dashed lines) and the fit to the data for the plant case (continuous lines) are also shown.
(b) Simulation (figures) and theoretical (lines) critical curves in the soil case for square (squares), triangular (triangles) and honeycomb (circles) lattices for several values of $I$: 0.0 (black), 0.1 (purple), 0.2 (green), 0.3 (cyan), 0.4 (blue) and 0.5 (red).} 
\label{fig:pb00}
\end{figure}

\begin{figure}
\centering
\includegraphics[scale=1]{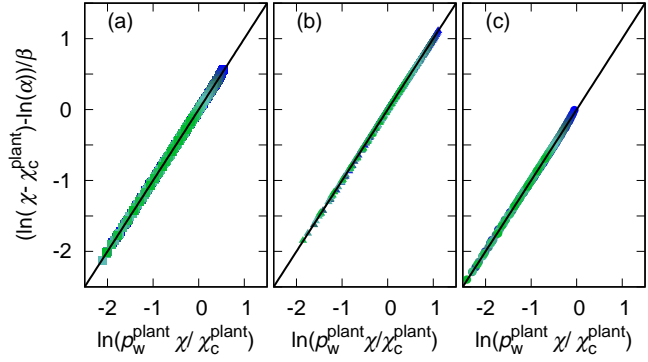}
\caption{Power law relation between $\chi$, $\chi_\text{c}^\text{plant}$ and $p_\text{w}^\text{plant}$ in the plant case when $I$ is fixed for a) square, b) triangular, and c) honeycomb lattices. Black solid line is the identity function. The color scale indicates the value of $I$ from $I=0$ (green) up to $I=1$ (blue) in steps of $\Delta I$=0.05.}
\label{fig:scaling}
\end{figure}

\begin{figure*}
\centering
\includegraphics[scale=1]{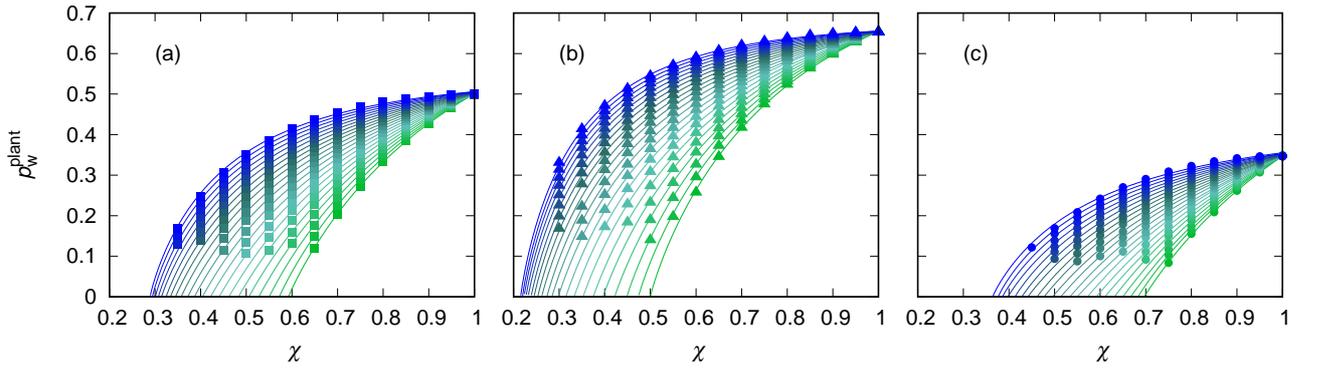}
\caption{Comparison between simulation results (figures) for $p_\text{w}^\text{plant}$ as a function of the susceptibility and the curve proposed in Eq.~\eqref{eq:pw-pants} (solid lines) for a) square, b) triangular and c) honeycomb lattices. The color scale indicates the value of $I$ from $I=0$ (green) up to $I=1$ (blue) in steps of $\Delta I$=0.05.}
\label{fig:pw-plants}
\end{figure*}

\section{Results}\label{results}
Simulation results for the critical curves of both soil and plant cases with no barriers are shown in Fig.~\ref{fig:pb00} a).
Notably, our results for $\chi_{c}^\text{soil}$ are very well described by the parametrization $p_\text{eff}^\text{soil}=I+(1-I)\chi=p_\text{cs}$.
Notice that the critical curves for $\chi_{c}^\text{plant}$ deviate from those for $\chi_{c}^\text{soil}$ for $I>0.15$.
This is due to non-susceptible plants lying in inoculated cells which do not belong to the clusters and can serve as a bridge between their adjacent sites.
We found that $\chi_{c}^\text{plant}$ can be well fitted by the Tsallis distribution $p_\text{cs}/(1+aI/n)^n$, with $a=$0.91$\pm$0.03 and 1.40$\pm$0.06 and $n=$2.0$\pm$0.4 and 1.1$\pm$0.1 for the square and triangular lattices, respectively.
For the honeycomb lattice $n$ takes a large value so we used $p_\text{cs}\exp(-aI)$ with $a=$0.63$\pm$0.01.
This behavior can be understood as the collective contribution of the interaction between susceptible plants and infected cells with a resistant plant. Note that the probability of observing this pair become higher as $\chi$ decreases and $I$ increases, and thus, the percolating system looks like a lattice formed by regular sites and sites involving complex nearest neighbors.
The main result of this analysis is the existence of a minimal susceptibility that guarantees the non-emergence of a spanning cluster of diseased plants even if all cells are inoculated, that is the value of $\chi_\text{c}^\text{plant}$ for $I=1$.
However, if $\chi>\chi_\text{c}^\text{soil}$ or $\chi>\chi_{c}^\text{plant}$ it is necessary to use the barrier strategy to reduce the connectedness of the lattice.
In Fig.~\ref{fig:pb00} b), we show the simulation results for the soil case.
Notice that they are well described by Eq.~\eqref{eq:pw-soil}, which corresponds to the description of the typical critical curves in the site-bond percolation with an occupation probability $p_\text{eff}^\text{soil}$.
This is because in this case the infected cells are taken into account in the cluster formation process even if the plant does not become sick.

On the other hand we found, for the plants case, that the relation between $\chi$, $\chi_\text{c}^\text{plant}$ and $p_\text{w}^\text{plant}$ is given by the power law  $(\chi-\chi_\text{c}^\text{plant})=\alpha(p_\text{w}^\text{plant}\chi/\chi_{c}^\text{plant})^\beta$ when $I$ is fixed, as it is shown in  Fig.~\ref{fig:scaling}.
It should be noted that both $\alpha$ and $\beta$
depend on $I$.
Particularly, $\beta$ takes values between 0.95 and 1.18 for all lattices.
Then, the critical curves for the plants case are given by
\begin{equation}
    p_\text{w}^\text{plant}=\frac{\chi_\text{c}^\text{plant}}{\chi}\left(\frac{\chi-\chi_\text{c}^\text{plant}}{\alpha} \right)^{1/\beta}
\label{eq:pw-pants}
\end{equation}
which matches very well the simulation data for the square, triangular and honeycomb lattices as shown in Fig.~\ref{fig:pw-plants} for different values of $I$. 
Table~\ref{tab:pfits} shows the values of the parameters $\alpha$ and $\beta$ (for different values of $I$) given by the fit to simulation data for the square, triangular and honeycomb lattices.
Moreover, in the case $\chi=1$, $p_\text{w}^\text{plant}=1-p_\text{cb}$ as expected since, under this condition, the system corresponds to the traditional bond percolation model.

\begin{table}[h]
\caption{Fit parameters for the square ($\Box$), triangular ($\triangle$) and honeycomb ($\bigcirc$) lattices. Error estimates in the last significant figure are indicated in parentheses.}
\label{tab:pfits}

\begin{center}

\begin{tabular}{c c c c c c c}

\hline
\hline

I&	$\alpha_\Box$&	$\beta_\Box$&	$\alpha_\triangle$&	$\beta_\triangle$&	$\alpha_\bigcirc$&	$\beta_\bigcirc$ \\

\hline
0.00&	0.4870(9)&	1.065(3)&	0.3685(5)&	1.132(5)&	0.621(5)&	1.031(8)\\
0.05&	0.4922(8)&	1.050(3)&	0.3689(5)&	1.099(5)&	0.637(2)&	1.032(3)\\
0.10&	0.4956(6)&	1.029(2)&	0.3673(4)&	1.077(3)&	0.636(2)&	1.001(3)\\
0.15&	0.4982(5)&	1.013(2)&	0.3646(4)&	1.051(3)&	0.657(3)&	1.003(5)\\
0.20&	0.5000(4)&	1.005(2)&	0.3598(4)&	1.032(3)&	0.669(3)&	0.996(5)\\
0.25&	0.4994(2)&	0.994(1)&	0.3530(2)&	1.029(1)&	0.674(2)&	0.974(4)\\
0.30&	0.4977(3)&	0.990(1)&	0.3445(1)&	1.024(1)&	0.685(2)&	0.974(3)\\
0.35&	0.4941(4)&	0.989(2)&	0.334(1)&	1.023(6)&	0.698(3)&	0.980(5)\\
0.40&	0.4892(3)&	0.991(2)&	0.3253(2)&	1.022(1)&	0.696(2)&	0.954(3)\\
0.45&	0.4821(3)&	0.996(2)&	0.3145(2)&	1.025(1)&	0.711(2)&	0.979(4)\\
0.50&	0.4741(2)&	1.003(1)&	0.3036(3)&	1.033(2)&	0.713(1)&	0.982(2)\\
0.55&	0.4653(2)&	1.013(1)&	0.2925(4)&	1.047(2)&	0.715(2)&	0.987(4)\\
0.60&	0.4551(2)&	1.025(1)&	0.2825(3)&	1.052(2)&	0.720(1)&	1.005(3)\\
0.65&	0.4450(1)&	1.036(1)&	0.2721(3)&	1.066(2)&	0.717(2)&	1.016(4)\\
0.70&	0.4342(3)&	1.050(2)&	0.2622(4)&	1.080(2)&	0.708(3)&	1.014(8)\\
0.75&	0.4235(5)&	1.070(3)&	0.2531(4)&	1.094(2)&	0.705(3)&	1.037(7)\\
0.80&	0.4120(6)&	1.082(4)&	0.2442(4)&	1.107(2)&	0.699(4)&	1.057(9)\\
0.85&	0.4017(8)&	1.104(5)&	0.2361(5)&	1.122(3)&	0.687(5)&	1.06(1)\\
0.90&	0.391(1)&	1.123(7)&	0.2285(5)&	1.137(2)&	0.676(5)&	1.08(2)\\
0.95&	0.380(1)&	1.142(8)&	0.2216(6)&	1.152(3)&	0.669(5)&	1.11(2)\\
1.00&	0.371(1)&	1.163(9)	&0.2148(7)&	1.165(4)&	0.654(6)&	1.12(2)\\
\hline
\hline
\end{tabular} 
\end{center}
\end{table}

\section{Application to chili plantations}\label{chili}

Application of Eq.~\eqref{eq:pw-pants} requires the knowledge of the plant's pathogen susceptibility. This quantity has been measured experimentaly as described in Ref.~\cite{jerc}. In general terms their method consists in sowing plants in previously sterilized soil and innoculating a fraction of the substrate with oomycetes. The pathogen is then allowed to propagate through the plantation and the presence of the pathogen is asessed for each plant. The ratio of the number of live infected plants to the total number of infected plants gives the surviving rate $\mathcal{P}$. The pathogen susceptibility of the plant is then calculated as $\chi=1-\mathcal{P}$.

\begin{figure}
\centering
\includegraphics[scale=1]{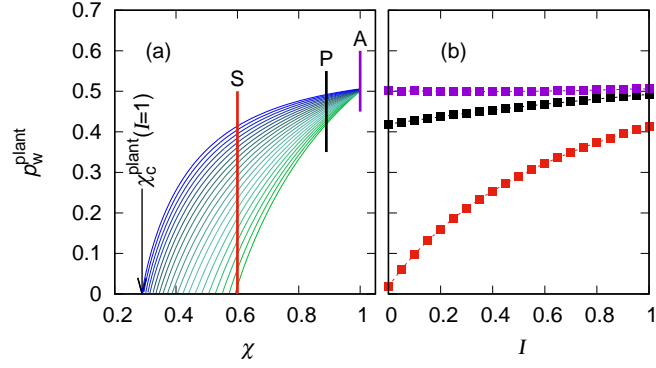}
\caption{(a) Critical values $p_\text{w}^\text{plant}$ for Arbol (A), Poblano (P) and Serrano (S) chili plants sowed with a sqare lattice arrangement. Vertical lines indicate their susceptibilities: 1.00 (A), 0.89 (P) and 0.60 (S). The solid curves are the same as in Fig.~\ref{fig:pw-plants}. $\chi_\text{c}^\text{plant}(I=1)=0.28883\pm0.00007$ is the maximum value of a plant's susceptibility that inhibits the formation of a cluster of diseased plants, even in the extreme case where the patogen is present all over the plantation.
(b) Values of $p_\text{w}^\text{plant}$ given by Eq.~\eqref{eq:pw-pants} and data from Table~\ref{tab:pfits} for the Arbol (purple), Poblano (black) and Serrano (red) chili plants on a square lattice.}
\label{fig:chiles}
\end{figure}

The reported values of the pathogen susceptibility for the varieties Arbol, Poblano and Serrano plants of chilis (which are of high commercial value un Mexico) are 1.00, 0.89 and 0.60, respectively. Putting these values into Eq.~\eqref{eq:pw-pants} we obtained the curves for $p_\text{w}^\text{plant}$ as a function of $I$ shown in Fig.~\ref{fig:chiles} for a square lattice. Note that as the value of $\chi$ approaches 1, like for the Arbol and Poblano chilis, the barrier density approaches the bond percolation threshold ($p_\text{cb}=0.5$) since in these particular cases the percolating system is very similar to the bond percolation model. On the other side, as $\chi$ approaches the site percolation threshold, like for the Serrano chili, the range of possible values for $p_\text{w}^\text{plant}$ becomes larger however $p_\text{w}^\text{plant}(I=1)\approx 0.41$ is less than 0.5. In practice this means an 18\% less barriers are needed to prevent the disease propagation.

Also, as $\chi$ becomes less and less than $p_\text{cs}$, the value of $p_\text{w}^\text{plant}$ decreases until it vanishes. This point, when $p_\text{w}^\text{plant}(I=1)=0$, corresponds to the intersection of the critical $\chi_\text{c}^\text{plant}$ curve with the vertical line $I=1$ (see Fig.~\ref{fig:pb00}). This is just the greatest value of a plant's susceptibility that makes the barrier strategy unnecessary.

\section{Conclusions}\label{conclusions}

In summary, we have presented a strategy based on the site-bond percolation model to prevent the propagation of \textit{Phytophthora} over a plantation. This strategy consists of placing barriers between adjacent cells, whose density depends on $\chi$ and $I$.
Two different clustering processes were analyzed: i) clusters of cells with the presence of the pathogen, and ii) clusters of diseased plants.
The former is related to a soil test and the latter to a direct visual inspection of the damage on the plantation.
It was found that both processes are indistinguishable, and therefore described by the same critical curve, for $I<0.15$.
On the contrary, for $I>0.15$ this behavior does not hold and different approaches for each process are necessary.
Differences in the critical density of barriers between the \emph{soil} and \emph{plant} cases are a consequence of the hybridization process of the lattice, which leads to a major deviation when $I$ increases and $\chi$ decreases (see Fig.~\ref{fig:pw-plants}).
The soil case is described by the site-bond percolation model with an effective occupation probability given by $p_\text{eff}^\text{soil}=I+(1-I)\chi$. Then the critical curves are as usual (see Eq.~\eqref{eq:pw-soil}) because the clustering process of the infected cells does not distinguish the sickness states of the plant.

In the plant case, the critical curves predict the existence of a minimal susceptibility $\chi_\text{c}^\text{plant}$ that guarantees a spanning cluster of infected plants will not appear, that is, if $\chi<\chi_\text{c}^\text{plant}$ even when $p_\text{w}=0$ and $I=1$.
Values for the minimal susceptibility in square, triangular and honeycomb lattices were found to be 0.28883$\pm$0.00007, 0.2141$\pm$0.0003 and 0.364$\pm$0.003, respectively.
Particularly, for the square lattice, this value is in agreement with the critical probability of lattices with more complex neighborhoods \cite{site-complex,site-complex2}.

Based on the obtained results, we would advise farmers and agronomists either to sow types of plants having a pathogen susceptibility lower than $\chi_\text{c}^\text{plant}$, or to apply the barriers strategy with a barrier density given by Eq.~\eqref{eq:pw-pants}. A very important advantage of this strategy is that it does not require to remove plants therefore avoiding deforestation.

This strategy could be verified under controlled conditions, for example, in greenhouses, tree nurseries, and hydroponics, where \textit{Phytophthora} and other phytopathogens cause great devastation.
On the other hand, its application on a real life situation requires to take into account other ecological and environmental variables, such as plant-plant or (beneficial) microorganism-plant interactions, irrigation system, spatial distribution of plants, the care provided by the farmer or the possibility of having more than one type of pathogen in the same parcel of soil.

Finally, Eq.~\eqref{eq:pw-pants} for $I=0$ could be used as an alternative parametrization of the critical curves in the site-bond percolation model even for lattices defined in dimensions higher than two.

J.\ E.\ R.\ acknowledges financial support from CONACyT (postdoctoral fellowship Grant no. 289198).
C.P. was supported by the grant Maria de Maeztu Unit of Excellence MDM-20-0692 and FPA Project No. 2017-83814-P of Ministerio de Ciencia, Innovación y Universidades (Spain), FEDER and Xunta de Galicia.


\begin{thebibliography}{10}

\bibitem{bio1}
D.~C. Erwin and O.~K. Ribeiro, {\em Phytophthora diseases worldwide.}
\newblock St. Paul, Minnesota, USA: American Phytopathological Society (APS
  Press), 1996.

\bibitem{bio2}
D.~Shaw, ``Phytophthora diseases worldwide'', The american phytopathological society
  (1996).
  {\em J. Agr. Sci.}, vol.~131, no.~2, pp.~245--249, 1998.

\bibitem{bio3}
T.~Reglinski, M.~Spiers, J.~Taylor, and M.~Dick, ``Root rot in radiata pine
  seedlings can be controlled.,'' {\em N. Z. J. Forestry}, vol.~54, no.~4,
  pp.~16--18, 2010.

\bibitem{chiles3}
F.~J. Polach and R.~K. Webster, ``Identification of strains and inheritance of
  pathogenicity in phytophthora capsici,'' {\em Phytopathology}, vol.~62,
  no.~1, pp.~20--26, 1972.

\bibitem{chiles}
M.~K. Hausbeck and K.~H. Lamour, ``Phytophthora capsici on vegetable crops:
  Research progress and management challenges,'' {\em Plant Dis.}, vol.~88,
  no.~12, pp.~1292--1303, 2004.

\bibitem{chiles2}
K.~H. Lamour, R.~Stam, J.~Jupe, and E.~Huitema, ``The oomycete broad-host-range
  pathogen phytophthora capsici,'' {\em Mol. Plant Pathol.}, vol.~13, no.~4,
  pp.~329--337, 2012.

\bibitem{papa}
H.~Lozoya-Salda{\~n}a, M.~N. Robledo-Esqueda, P.~Rivas-Valencia,
  S.~Sandoval-Islas, M.~T. Beryl Colinas~y Le{\'o}n, and C.~Nava-D{\'\i}az,
  ``Sensitivity to fungicides of <em>phytophthora infestans</em> (mont.) de
  bary in chapingo, mexico,'' {\em Rev. Chapingo Ser. Horticultura}, vol.~23,
  no.~3, pp.~175--186, 2017.

\bibitem{papa2}
S.~K. Shakya, M.~M. Larsen, M.~M. Cuenca-Condoy, H.~Lozoya-Salda{\~n}a, and
  N.~J. Gr{\"u}nwald, ``Variation in genetic diversity of phytophthora
  infestans populations in mexico from the center of origin outwards,'' {\em
  Plant Dis.}, vol.~102, no.~8, pp.~1534--1540, 2018.

\bibitem{papa3}
B.~J. Haas {\em et~al.}, ``Genome sequence and analysis of the irish potato
  famine pathogen phytophthora infestans,'' {\em Nature}, vol.~461, p.~393, 09
  2009.

\bibitem{aguacate3}
G.~Weste and G.~C. Marks, ``The biology of phytophthora cinnamomi in
  australasian forests,'' {\em Annu. Rev. Phytopathol.}, vol.~25, no.~1,
  pp.~207--229, 1987.

\bibitem{aguacate}
M.~You and K.~Sivasithamparam, ``Changes in microbial populations of an avocado
  plantation mulch suppressive of phytophthora cinnamomi,'' {\em Appl. Soil
  Ecology}, vol.~2, no.~1, pp.~33 -- 43, 1995.

\bibitem{aguacate2}
H.~S. Judelson and B.~Messenger-Routh, ``Quantitation of phytophthora cinnamomi
  in avocado roots using a species-specific dna probe,'' {\em Phytopathology},
  vol.~86, no.~7, pp.~763--768, 1996.

\bibitem{ink1}
A.~Vannini and A.~M. Vettraino, ``Ink disease in chestnuts: impact on the
  european chestnut,'' {\em For. Snow Landsc. Res.}, vol.~76, no.~3,
  pp.~345--350, 2001.

\bibitem{ink2}
A.~M. Vettraino, O.~Morel, C.~Perlerou, C.~Robin, S.~Diamandis, and A.~Vannini,
  ``Occurrence and distribution of phytophthora species in european chestnut
  stands, and their association with ink disease and crown decline,'' {\em Eur.
  J. Plant Pathol.}, vol.~111, no.~2, p.~169, 2005.

\bibitem{ink3}
A.~Vannini, G.~Natili, N.~Anselmi, A.~Montaghi, and A.~M. Vettraino,
  ``Distribution and gradient analysis of ink disease in chestnut forests,''
  {\em Forest Pathol.}, vol.~40, no.~2, pp.~73--86, 2010.

\bibitem{castano}
E.~Vieitez, ``El casta{\~n}o y sus procesos de rizog{\'e}nesis,'' {\em Trabajos
  Dept. Bot. Fisiol. Veg}, vol.~11, pp.~25--31, 1981.

\bibitem{castano2}
M.~V. Gonz{\'a}lez, B.~Cuenca, M.~L{\'o}pez, M.~J. Prado, and M.~Rey,
  ``Molecular characterization of chestnut plants selected for putative
  resistance to phytophthora cinnamomi using ssr markers,'' {\em Sci. Hortic.},
  vol.~130, no.~2, pp.~459 -- 467, 2011.

\bibitem{castano3}
M.~Miranda-Fontaina, J.~Fern{\'a}ndez-L{\'o}pez, A.~Vettraino, and A.~Vannini,
  ``Resistance of castanea clones to phytophthora cinnamomi: testing and
  genetic control,'' {\em Silvae Genet.}, vol.~56, no.~1-6, pp.~11--21, 2007.

\bibitem{bio6}
D.~C. Erwin and O.~K. Ribeiro, {\em Phytophthora diseases worldwide.}
\newblock American Phytopathological Society (APS Press), 1996.

\bibitem{bio7}
E.~Bernhardt and R.~Grogan, ``Effect of soil matric potential on the formation
  and indirect germination of sporangia of \textit{Phytophthora parasitica},
  \textit{Phytophthora capsici}, and \textit{Phytophthora capsici cryptogea}
  rots of tomatoes, \textit{Lycopersicon esculentum},'' {\em Phytopathology
  (USA)}, vol.~72, no.~5, 1982.

\bibitem{bio8}
A.~R. Hardham, ``\textit{Phytophthora cinnamomi},'' {\em Mol. Plant Pathol.},
  vol.~6, no.~6, pp.~589--604, 2005.

\bibitem{bio9}
B.~Feng, P.~Li, H.~Wang, and X.~Zhang, ``Functional analysis of pcpme6 from
  oomycete plant pathogen \textit{Phytophthora capsici},'' {\em Microb.
  Pathogenesis}, vol.~49, no.~1, pp.~23 -- 31, 2010.

\bibitem{bio10}
P.~Li, B.~Feng, H.~Wang, P.~W. Tooley, and X.~Zhang, ``Isolation of nine
  \textit{Phytophthora capsici} pectin methylesterase genes which are
  differentially expressed in various plant species,'' {\em J. Basic Microb.},
  vol.~51, no.~1, pp.~61--70, 2011.

\bibitem{fungi3}
Y.~Cohen and M.~D. Coffey, ``Systemic fungicides and the control of
  oomycetes,'' {\em Annual Review of Phytopathology}, vol.~24, no.~1,
  pp.~311--338, 1986.

\bibitem{fungi2}
A.~Drenth and D.~I. Gest, ``\textit{Phytophthora} in the tropics,'' in {\em
  Diversity and Management of \textit{Phytophthora} in Southeast Asia}
  (A.~Drenth and D.~I. Gest, eds.), no.~114, ch.~Biology of
  \textit{Phytophthora}, pp.~30--41, ACIAR, 2004.

\bibitem{sp}
H.~L. Frisch and J.~M. Hammersley, ``Percolation processes and related
  topics,'' {\em J. Soc. Ind. Appl. Math.}, vol.~11, no.~4, pp.~894--918, 1963.

\bibitem{sp4}
D.~S. Callaway, M.~E.~J. Newman, S.~H. Strogatz, and D.~J. Watts, ``Network
  robustness and fragility: Percolation on random graphs,'' {\em Phys. Rev.
  Lett.}, vol.~85, pp.~5468--5471, Dec 2000.

\bibitem{sp2}
M.~E.~J. Newman, ``Spread of epidemic disease on networks,'' {\em Phys. Rev.
  E}, vol.~66, p.~016128, Jul 2002.

\bibitem{sp3}
N.~Madar, T.~Kalisky, R.~Cohen, D.~ben Avraham, and S.~Havlin, ``Immunization
  and epidemic dynamics in complex networks,'' {\em Eur. Phys. J. B}, vol.~38,
  pp.~269--276, Mar 2004.

\bibitem{barrier1}
W.~A. Dunstan, T.~Rudman, B.~L. Shearer, N.~A. Moore, T.~Paap, M.~C. Calver,
  R.~Armistead, M.~P. Dobrowolski, B.~Morrison, K.~Howard, E.~O'Gara, C.~Crane,
  B.~Dell, P.~O'Brien, J.~A. McComb, and G.~E. S.~J. Hardy, ``Research into
  natural and induced resistance in australian native vegetation of
  phytophthora cinnamomi and innovative methods to contain and/or eradicate
  within localised incursions in areas of high biodiversity in australia.,''
  Tech. Rep. 19/2005DEH, Centre for Phytophthora Science and Management for the
  Australian Government Department of the Environment, Water, Heritage and the
  Arts, 2008.

\bibitem{barrier2}
C.~P. Dunne, C.~E. Crane, M.~Lee, T.~Massenbauer, S.~Barrett, S.~Comer, G.~J.
  Freebury, D.~J. Utber, M.~J. Grant, and B.~L. Shearer, ``A review of the
  catchment approach techniques used to manage a phytophthora cinnamomi
  infestation of native plant communities of the fitzgerald river national park
  on the south coast of western australia,'' {\em N. Z. J. Forestry Sci.},
  vol.~41, p.~S121, 2011.

\bibitem{barrier3}
W.~A. Dunstan, T.~Rudman, B.~L. Shearer, N.~A. Moore, T.~Paap, M.~C. Calver,
  B.~Dell, and G.~E. S.~J. Hardy, ``Containment and spot eradication of a
  highly destructive, invasive plant pathogen (phytophthora cinnamomi) in
  natural ecosystems,'' {\em Biol. Invasions}, vol.~12, p.~913, 2011.

\bibitem{barrier4}
B.~L. Shearer, C.~E. Crane, R.~G. Fairman, M.~J. Dillon, and R.~M. Buehrig,
  ``Spatio-temporal variation in invasion of woodlands and forest by
  phytophthora cinnamomi,'' {\em Australas. Plant Path.}, vol.~43,
  pp.~327--337, May 2014.

\bibitem{edu1}
G.~Bonanomi, V.~Antignani, C.~Pane, and F.~Scala, ``Suppression of soilborne
  fungal diseases with organic amendments,'' {\em J. Plant Pathol.}, vol.~89,
  no.~3, pp.~311--324, 2007.

\bibitem{edu2}
H.~A.~J. Hoitink, L.~V. Madden, and A.~E. Dorrance, ``Systemic resistance
  induced by trichoderma spp.: Interactions between the host, the pathogen, the
  biocontrol agent, and soil organic matter quality,'' {\em Phytopathology},
  vol.~96, no.~2, pp.~186--189, 2006.

\bibitem{edu3}
C.~M. Craft and E.~B. Nelson, ``Microbial properties of composts that suppress
  damping-off and root rot of creeping bentgrass caused by pythium
  graminicola.,'' {\em Appl. Environ. Microb.}, vol.~62, no.~5, pp.~1550--1557,
  1996.

\bibitem{walk}
C.~E.~M. van, den~Boom, T.~A. van Beek, and M.~Dicke, ``Differences among plant
  species in acceptance by the spider mite tetranychus urticae koch,'' {\em J.
  Appl. Entomol.}, vol.~127, no.~3, pp.~177--183, 2003.

\bibitem{splash}
L.~Geagea, L.~Huber, I.~Sache, D.~Flura, H.~McCartney, and B.~Fitt, ``Influence
  of simulated rain on dispersal of rust spores from infected wheat
  seedlings,'' {\em Agr. Forest Meteorol.}, vol.~101, no.~1, pp.~53 -- 66,
  2000.

\bibitem{splash3}
S.~Saint-Jean, A.~Testa, L.~Madden, and L.~Huber, ``Relationship between
  pathogen splash dispersal gradient and weber number of impacting drops,''
  {\em Agr. Forest Meteorol.}, vol.~141, no.~2, pp.~257 -- 262, 2006.

\bibitem{splash2}
T.~Gilet and L.~Bourouiba, ``{Rain-induced Ejection of Pathogens from Leaves:
  Revisiting the Hypothesis of Splash-on-Film using High-speed
  Visualization},'' {\em Integr. Comp. Biol.}, vol.~54, pp.~974--984, 10 2014.

\bibitem{pythium}
C.~A. Walker and P.~van West, ``Zoospore development in the oomycetes,'' {\em
  Fungal Biol. Rev.}, vol.~21, no.~1, pp.~10 -- 18, 2007.

\bibitem{jerc}
J.~E. Ram\'{\i}rez, E.~Molina-Gayosso, J.~Lozada-Lechuga, L.~M. Flores-Rojas,
  M.~I. Mart\'{\i}nez, and A.~Fern\'andez~T\'ellez, ``Percolation strategy to
  improve the production of plants with high pathogen susceptibility,'' {\em
  Phys. Rev. E}, vol.~98, p.~062409, Dec 2018.

\bibitem{tarasevich}
Y.~Y. Tarasevich and S.~C. van~der Marck, ``An investigation of site-bond
  percolation on many lattices,'' {\em Int. J. Mod. Phys. C}, vol.~10, no.~07,
  pp.~1193--1204, 1999.

\bibitem{Ziff1}
M.~E.~J. Newman and R.~M. Ziff, ``Efficient monte carlo algorithm and
  high-precision results for percolation,'' {\em Phys. Rev. Lett.}, vol.~85,
  pp.~4104--4107, Nov 2000.

\bibitem{Ziff2}
M.~E.~J. Newman and R.~M. Ziff, ``Fast monte carlo algorithm for site or bond
  percolation,'' {\em Phys. Rev. E}, vol.~64, p.~016706, Jun 2001.

\bibitem{Mertens}
S.~Mertens and C.~Moore, ``Continuum percolation thresholds in two
  dimensions,'' {\em Phys. Rev. E}, vol.~86, p.~061109, Dec 2012.

\bibitem{Rintoul_1997}
M.~D. Rintoul and S.~Torquato, ``Precise determination of the critical
  threshold and exponents in a three-dimensional continuum percolation model,''
  {\em J. Phys. A-Math. Gen.}, vol.~30, pp.~L585--L592, aug 1997.

\bibitem{stauffer}
D.~Stauffer and A.~Aharony, {\em Introduction to percolation theory}.
\newblock Taylor \& Francis, 2014.

\bibitem{Coniglio}
A.~Coniglio, ``Cluster structure near the percolation threshold,'' {\em J.
  Phys. A-Math. Gen.}, vol.~15, no.~12, pp.~3829--3844, 1982.

\bibitem{site-complex}
K.~Malarz and S.~Galam, ``Square-lattice site percolation at increasing ranges
  of neighbor bonds,'' {\em Phys. Rev. E}, vol.~71, p.~016125, Jan 2005.

\bibitem{site-complex2}
M.~Majewski and K.~Malarz, ``Square lattice site percolation thresholds for
  complex neighbourhoods,'' {\em Acta Phys. Pol. B}, vol.~38, p.~2191, Jun
  2007.

\end{thebibliography}
\end{document}